\documentclass[12pt,preprint]{aastex}
\usepackage{emulateapj5,epsf}

\shortauthors{T. Hosokawa \& S.Inutsuka}
\shorttitle{A Mode of Triggered Star Formation}

\begin{document}

\title{Dynamical Expansion of Ionization and Dissociation Front around
a Massive Star. I. A Mode of Triggered Star Formation}
\author{Takashi Hosokawa\altaffilmark{1} and 
     Shu-ichiro Inutsuka\altaffilmark{2}}

\altaffiltext{1}{Yukawa Institute for Theoretical Physics, Kyoto University,
Kyoto, Japan, 606-8502 ; hosokawa@yukawa.kyoto-u.ac.jp}
\altaffiltext{2}{Department of Physics, Kyoto University, Kyoto, Japan, 
606-8502 ; inutsuka@tap.scphys.kyoto-u.ac.jp} 

\begin{abstract}
We analyze the dynamical expansion of the H~II region and outer
photodissociation region (PDR) around a massive star by solving
the UV and FUV radiation transfer and the thermal and chemical
processes in a time-dependent hydrodynamics code. 
We focus on the physical structure of the shell swept up by the shock
front (SF) preceding the ionization front (IF). 
After the IF reaches the initial Str$\ddot {\rm o}$mgren radius, 
the SF emerges in front of the IF and the geometrically
thin shell bounded with the IF and the SF is formed.
The gas density inside the shell is about $10^{1-2}$ times 
as high as the ambient gas density.
Initially the dissociation fronts expands faster than IF and the PDR
is formed outside the H~II region. Thereafter the IF and SF
gradually overtakes the proceeding dissociation fronts (DFs),  and 
eventually DFs are taken in the shell.  
The chemical composition within the shell is initially
atomic, but hydrogen and carbon monoxide molecules are gradually
formed. This is partly because the IF and SF overtake DFs and SF
enters the molecular region, and partly because the reformation
timescales of the molecules become shorter than the dynamical timescale.  
The gas shell becomes dominated by the molecular gas by the
time of gravitational fragmentation, which agrees with some recent 
observations. A simple estimation of star formation rate in the shell 
can provide a significant star formation rate in our galaxy.
\end{abstract}
\keywords{ Circumstellar matter -- H~II regions -- ISM: molecules 
           -- STARS : formation}

\section{Introduction}

As the H~II region expands around a massive star, the shock front (SF)
emerges preceding the ionization front (IF). This SF sweeps up the
ambient gas and the gas shell is formed at the edge of the H~II
region. Many authors have studied the scenario that 
the shell becomes unstable and next star formation is triggered
around H~II region (collect and collapse model, see, e.g., 
Elmegreen \& Lada 1977, Whitworth et al. 1994, Elmegreen 1998,
and references therein). 
Recently, Deharveng et al.(2003, hereafter DV03) observed the 
fragmented molecular shell around the classical H~II region, 
Sh104 and show the young stars (cluster) is formed in the 
core of one fragment. They argue that this is the evidence 
of collect and collapse model. Now we can refine the theory for 
the evolution of the shell and compare it with the detailed observation.  
Especially, we can focus on the amount of the fragmented {\it molecular} 
gas they observed. The SF initially emerges in front of the IF, where 
sufficient FUV photons to dissociate molecules are available.
Therefore, it is not clear whether the SF can gather the molecular 
gas in the shell as observation indicates or not.

Roger \& Dewdney (1992) and Diaz-Miller, Franco
\& Shore (1998) studied the time-dependent expansion of H~II and 
PDR solving the radiative transfer of UV and FUV photons. 
Although these works do not include hydrodynamics, they have shown 
that the IF gradually overtakes the dissociation front (DF) and 
this means that IF and preceding SF gradually enter the molecular region. 
The numerical study of the hydrodynamical expansion 
of the IF has been studied since 1960's 
(e.g. Mathews 1965, Lasker 1966, Tenorio-Tagle 1976, Tenorio-Tagle 1979 
, Franco et al. 1990 etc.). These works successfully show the various
dynamical aspects of the expansion in both homogeneous and
inhomogeneous ambient medium. However, these do not include
the outer photodissociation region (PDR) and thermal processes dominant in PDR.

We perform the time-dependent calculation including the IF, 
DF, and the shell, which has been very limited.
We solve the UV and FUV radiation transfer and hydrodynamics 
numerically to investigate the structure and evolution 
of the shell and PDR as well as H~II region. 
Since the compression rate behind the SF depends on the 
thermal processes, we include the thermal processes which dominate 
either in H~II region or PDR. In this letter, we consider the simple 
situation that there is one massive star within the homogeneous 
ambient medium, which can be compared with Sh104 observed by DV03. 
The detailed quantitative aspects of the evolution
will be given in subsequent paper (Hosokawa \& Inutsuka 2004,
hereafter, Paper II).

\section{Numerical Methods}

We use a one-dimensional spherically-symmetrical numerical method.
The numerical scheme for the hydrodynamics is based on the 2nd-order 
Lagrangian Godunov method (see, e.g., van Leer 1979).
We use the on-the-spot approximation for UV and FUV radiation transfer. 
We assume that all hydrogen molecules are in the ground vibrational
level of X$^1 \Sigma_{\rm g}$ and ortho/para ratio is 3:1
(Diaz-Miller, Franco \& Shore 1998).
For the Lyman-Werner bands to photodissociate hydrogen molecule,
we solve the frequency dependent transfer using the representative 
set of lines. 
The UV and FUV photon luminosity of the central star is adopted from
Diaz-Miller, Franco \& Shore (1998) for the case of $Z = 1 Z_\odot$.  
The main thermal processes included in the energy equation is UV/FUV 
heating (e.g., H photoionization, photoelectric heating) and radiative 
cooling (e.g., H recombination, Ly-$\alpha$, OI~(63.0$\mu$m, 63.1$\mu$m)
, OII~(37.29$\mu$m), CII~(23.26$\mu$m, 157.7$\mu$m), H$_2$, and CO,
see Hollenbach \& McKee 1979, 1989, Koyama \& Inutsuka 2000). 
Non-equilibrium reaction equations are implicitly solved for the 
species of e, H, H$^+$, H$_2$, C$^+$, and CO. 
The ionization rate of O is assumed to be the same as that of
H. We adopt the simple approximation for the dissociation process 
of CO molecule given by Nelson \& Langer (1997).  
The hydrodynamic equations and the radiative transfer equation, 
energy equation, and reaction equations are combined following 
Tenorio-Tagle (1976).
The dust extinction is included only outside the IF and the dust
temperature is calculated following Hollenbach, Takahashi \& Tielens
(1991). 
We have checked our numerical code with the well-studied problems 
(e.g., Franco et al. 1990).

\section{Results of the Numerical Calculation}

In this letter, we consider one typical case where there is 
one massive star of $T_{\rm eff} =40000~{\rm K}~( M=41~M_\odot )$ 
in the ambient medium. 
This corresponds to the O6V star, which is the
central star of Sh104.  The initial ambient number density of the hydrogen
nucleon is unknown, and we adopt a typical value of the giant
molecular cloud,  $n_{\rm H,0} = 10^3~{\rm cm}^{-3}$ 
$( n_{\rm H_2,0} = 500~{\rm cm}^{-3} )$.
As in the standard picture (e.g. Spitzer 1978),
the IF expands rapidly as the weak R-type front in the recombination
timescale, $t_{\rm rec} \sim 100 $~yr for $n_{\rm H,0} = 10^3~{\rm cm}^{-3}$ 
(formation phase).  
As the IF reaches Str$\ddot {\rm o}$mgren radius, $R_{\rm st} = 0.56$~pc,
the photoionization rate and the recombination rate in the
H~II region becomes equal, the H~II region begins to expand owing to
the pressure difference between H~II region and outer PDR or molecular
cloud. At this phase, the SF appears in front of the
IF and the IF changes to the weak D-type front (expansion phase).
The expansion law in this phase is given by

\begin{equation}
R_{\rm IF}(t) = R_{\rm st} \left( 1 + \frac74 
                                      \frac{C_{\rm HII}t}{R_{\rm st}}
                           \right)^{4/7}
\label{law}
\end{equation}
(e.g. Spitzer 1978), where $C_{\rm HII}$ is the sound speed 
in the H~II region.
 The dynamical timescale is now given by 
$t_{\rm dyn} = R_{\rm st}/C_{\rm HII} \sim 10^5$~yr. 
Fig.1 shows the time evolution of the physical quantities 
of the gas in the expansion phase.
The radius of Sh104 is 4~pc and the IF reaches this radius at
 $t=0.7$~Myr in our calculation.
For each snapshot, we can see the H~II region $( T \sim 10^4~{\rm K} )$,
the gas shell swept up by the SF 
$( n_{\rm H} \sim 10^{4-5}~{\rm cm^{-3}} )$ and the outer PDR 
$( T \sim 100~{\rm K} )$ and the outermost molecular cloud 
$( T \sim 10~{\rm K} )$.
The SF emerges just in front of the IF. 
The density behind the SF significantly increases from the ambient 
value owing to the line cooling, and the gas density 
within the shell is about 10-100 times as high as the ambient density.
The more detailed inner structure of the shell is shown below (Fig.3).
The geometrical thickness of the shell, $h$ increases throughout the
calculation, $h \sim 0.02-0.2~{\rm pc}$. 
The PDR $( T \sim 100~{\rm K} )$ becomes narrow as the IF
expands, that is, the IF gradually catches 

\vspace{3mm}
\begin{center}
\epsfxsize=8cm 
\epsfbox{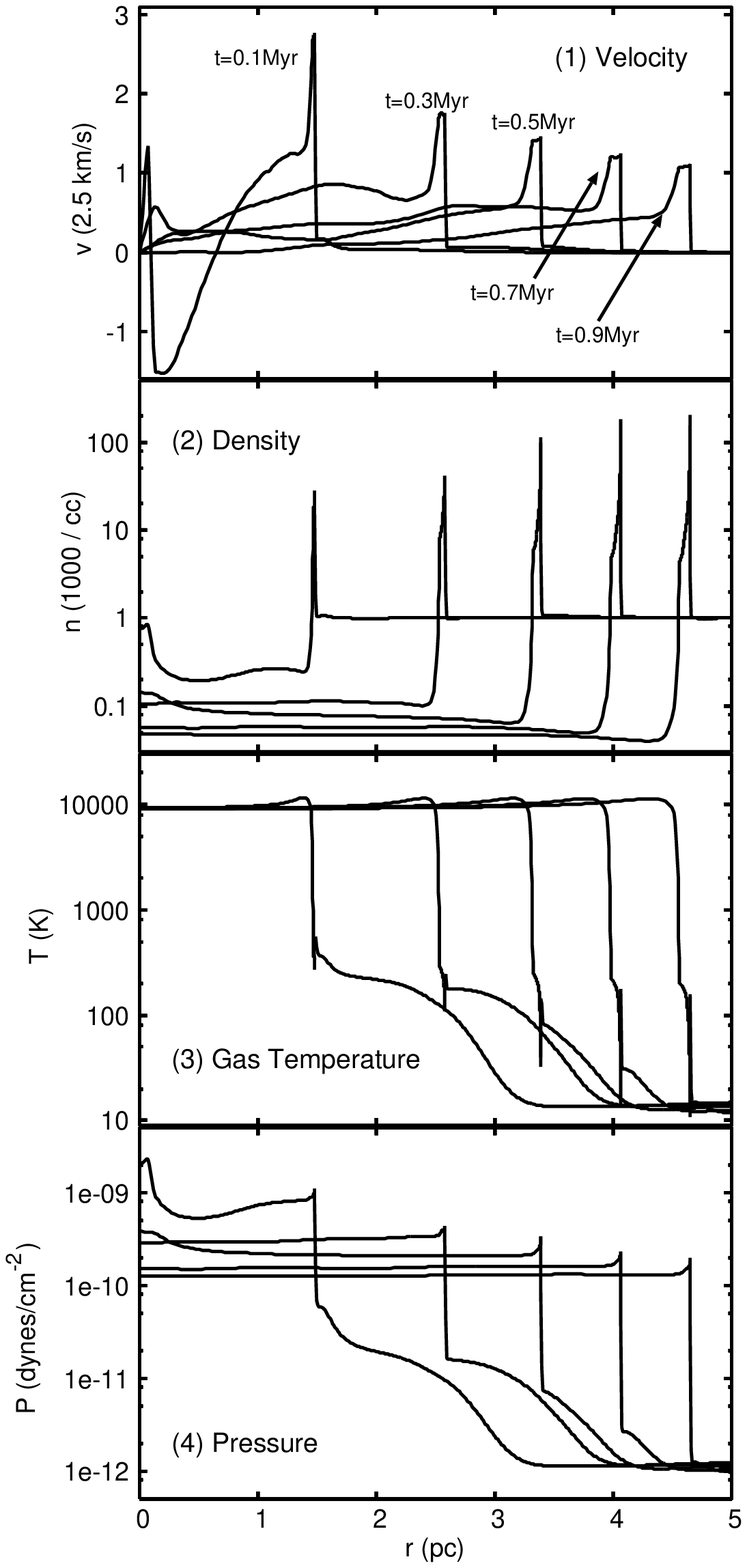}
\figcaption{The snapshots of the gas-dynamical evolution. 
In each panel, five snapshots represents the profiles at 
$t = 0.1, 0.3, 0.5, 0.7$, and 0.9~Myr respectively.}
\end{center}
\vspace{-4mm}

\noindent
up with the preceding DFs 
(e.g. Roger \& Dewdney 1992,  Diaz-Miller, Franco \& Shore 1998).
The upper panel of Fig.2 represents the overtaking of the IF more
clearly. The position of the SF is very close to 
that of the IF.
As the system switches from the formation phase to the 
expansion phase, the timescale of the expansion of the IF suddenly 
becomes longer to $t_{\rm dyn}$.
Even in this phase, the preceding DFs continue to expand. Then
the region between IF and DFs appears as a PDR. 
However, the DFs gradually slow down because the FUV flux at each DF 
decreases owing to the geometrical dilution and the dust extinction in the
 PDR.  The IF expands because of the pressure difference between the
 H~II region and the outer PDR, and the deceleration is more slower 
according to eq.(\ref{law}).

\vspace{3mm}
\begin{center}
\epsfxsize=8cm 
\epsfbox{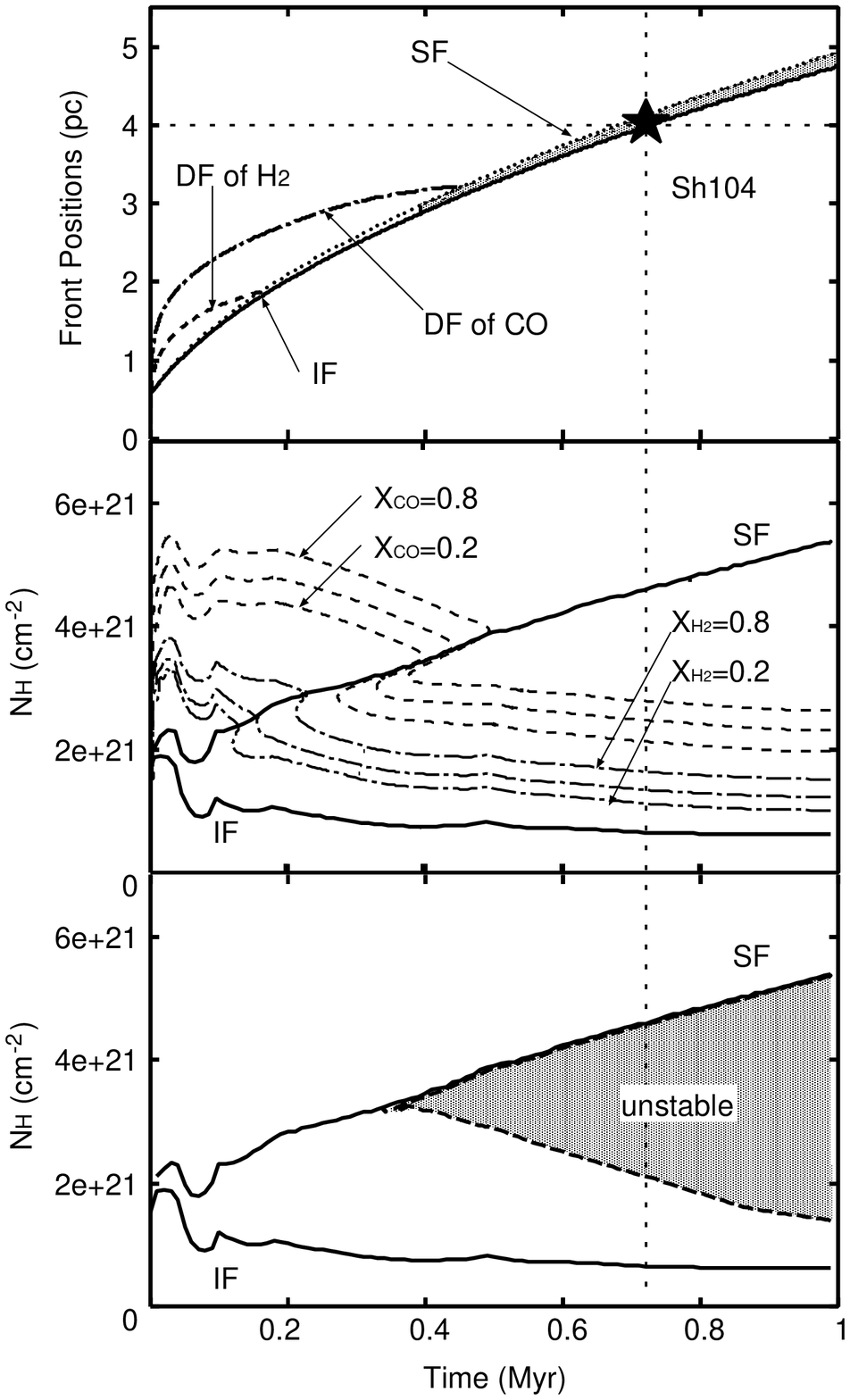}
\figcaption{
{\it Upper panel} : The time evolution of the various front positions. 
The solid and dotted lines mean the position of the IF and SF respectively. 
The broken (dot-solid) line represents the DF of H$_2$ (CO).
To define the DFs of H$_2$ and CO, we use 
${\rm X_{\rm H_2}} \equiv 2 n_{\rm H_2}/n_{\rm H_{\rm nuc}}=0.5$
and $ {\rm X_{\rm CO}} \equiv n_{\rm CO}/n_{\rm C_{\rm nuc}} = 0.5$.
The shaded region corresponds to $t < (G \rho)^{-1/2}$ within the
shell, and hence, the gravitational fragmentation is expected (see \S4.1).
{\it Middle panel} : The time evolution of the column density of each region.
Thin contour lines represent the position where 
${\rm X}_{\rm H_2}$ (broken), ${\rm X}_{\rm CO}$ (dot-solid)
= 0.2, 0.5 and 0.8.
{\it Lower panel} : Same as the meddle panel but for showing
the gravitationally unstalbe region.
The shaded region corresponds to the region where $t < (G \rho)^{-1/2}$. 
}
\end{center}

\noindent
The lower two panels of Fig.2 represents the time evolution of the
column density of each region. The column density of the H~II region
gradually decreases because the number density decreases as 
$n_{\rm HII} \propto R_{\rm IF}^{-3/2}$ in the expansion phase, where
$R_{\rm IF}$ is the radius of the H~II region. The column density
of the gas shell $\sigma$ increases.  
The ambient mass swept up by the time $t$ is proportional to 
$R_{\rm IF}(t)^3$ and the mass within the H~II region is 
$\propto n_{\rm HII} R_{\rm IF}(t)^3 \propto R_{\rm IF}(t)^{3/2}$.
Then the most of the swept-up mass does not flow into the H~II region but 
remains in the shell. The calculated mass of the H~II region is 
$430 M_\odot$ and the shell mass is $9000 M_\odot$ at $t=0.7$~Myr
(including both PDR and molecular region), which is in good agreement
with the observation (see DV03). 
The PDR initially spreads outside the shell, but H$_2$ and CO
molecules are gradually accumulated from the outer side of the shell.
The accumulation of the molecules already begins when each DF is still 
outside the
 
\vspace{5mm}
\epsfxsize=8.0cm 
\epsfbox{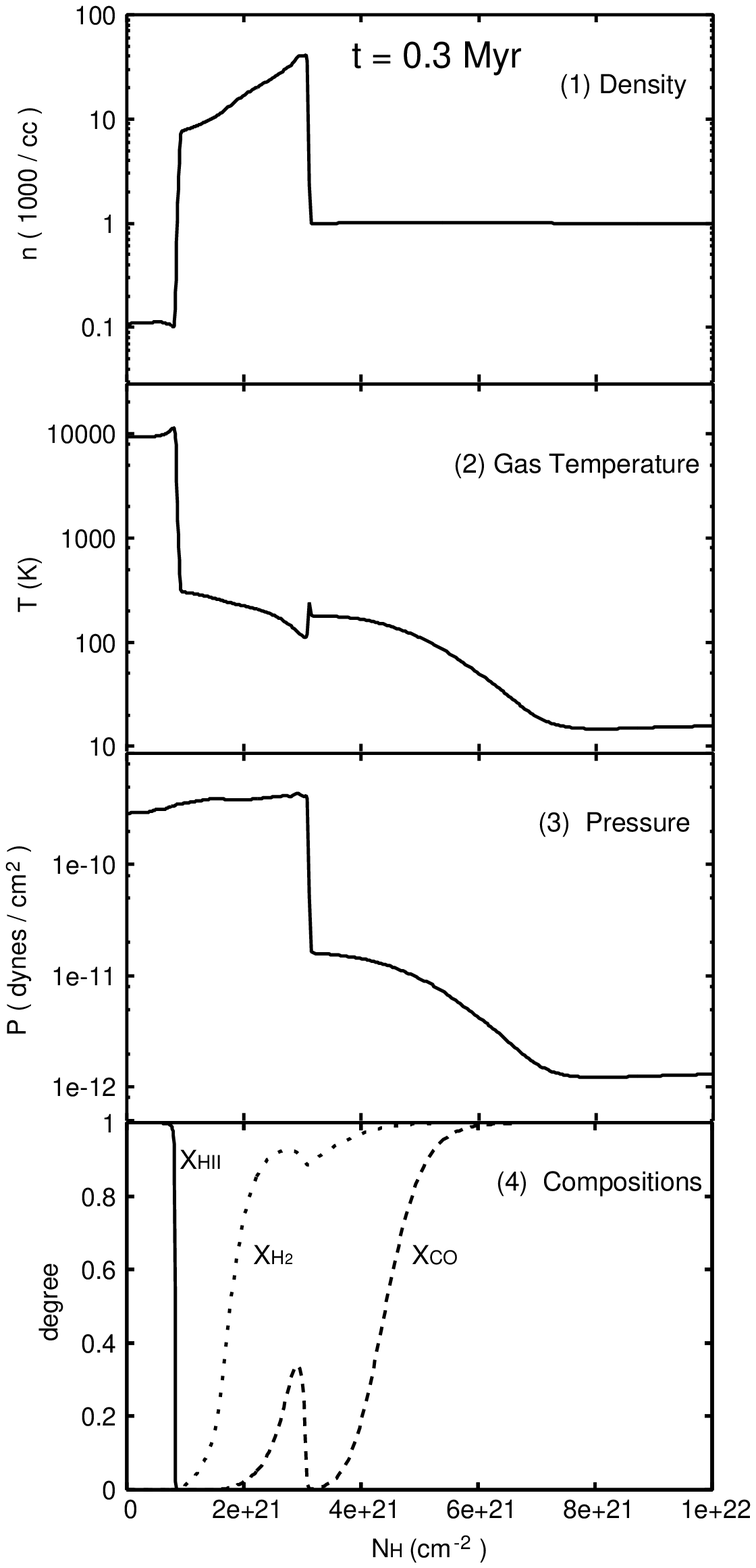}
\figcaption{
The distribution of gas density, pressure,
temperature, and the chemical compositions at $t=0.3$~Myr.
In the lowest panel, ${\rm X}_{\rm H_2}$ and ${\rm X}_{\rm CO}$ are
molecular ratios (see the caption of Fig.2) and ${\rm X}_{\rm HII}$
is the ionization rate defined as $n_{\rm H^+}/n_{\rm H_{\rm nuc}}$. }
\vspace{8mm}

\noindent
SF.
Therefore, the accumulation is not simply explained by 
the fact that the IF and the SF overtakes the DFs and that the SF
enters the molecular region.
The gas density within the shell is so high that the 
reformation timescales of the molecules become short enough.
The reformation timescale of the hydrogen molecule is about 
$\sim 10^6$yr for $n_{\rm H,0} \sim 10^3 {\rm cm^{-3}}$. 
The dynamical timescale is about $t_{\rm dyn} \sim 10^5$~yr, 
then the reformation of H$_2$ on the grain surface is not 
important with the ambient  number density. However, the reformation 
timescale in the gas shell, $\sim 10^{4-5}$~yr, is shorter
than $t_{\rm dyn}$,
therefore the formation of H$_2$ molecule is significant in the shell.

Fig.3 shows the detailed structure of the physical quantities and the
chemical composition of the shell at $t=0.3$~Myr. 
As this figure shows, the molecular fraction in the shell is higher 
than the value in the PDR outside the shell owing to the rapid reformation.  
As H~II region expands, the DFs are taken in the shell and the denser 
($n_{\rm H} > 10^4  {\rm cm^{-3}}$) PDR is formed in the inner side 
of the shell.
Once the molecules are reformed in the shell, the FUV radiation is
consumed by the photodissociation of these molecules, which decelerates the 
expansion of the preceding DFs to accelerates the accumulation
of the molecules. 
About 80 \% (55 \%) of the hydrogen (carbon) atoms within the shell
exist as H$_2$ (CO) molecule at $t = 0.7$~Myr in our calculation.
Fig.3 also shows other interesting features of the shell.
The thermal pressure inside the shell is high and about constant 
at the value of the H~II region. 
The gas temperature has negative gradient from the inner side to the
outer side of the shell, because the FUV radiation is attenuated owing to the
dust extinction through the shell, and the photoelectric heating
decreases.  Conversely, the gas density has positive gradient to the
outer side of the shell.

\section{Discussion}

\subsection{Dust in H~II Region}

In this section, we discuss the possible effect of dust grains
in the H~II regions (e.g. Spitzer 1978).
For example, the grains can be driven by the radiation pressure 
during the expansion of the H~II region (e.g., Arthur et al. 2004).
Arthur et al. have shown the small grains are destroyed by the sublimation
but the large graphite grains can exist close to the star. 
In contrast, the recent observation by Roger et al. (2004) found no 
far-infrared component correlated with H~II region, Sh170. 
The photoelectric heating works even in H~II regions comparatively 
with the photoionization heating (Weingartner \& Draine 2001)
and is sensitive to the uncertain abundance of the small dust grains. 
If we adopt the dust absorption cross section,
$6 \times 10^{-22} {\rm cm}^2$ (H nucleon)$^{-1}$ in H~II region, 
which we adopt in the PDR and the outer region, the radius of the 
H~II region becomes slightly smaller, but the difference from the 
case without dust in H~II region is always less than 5\%. 
The temperature in the PDR decreases by 10-20\% because of the 
attenuation of FUV photons by dust grains in the H~II region.
The gas density within the shell rises by a factor, 1.4-1.5.
Theses changes in the shell, however, only slightly 
promotes the accumulation of the molecular gas.
If we include the standard photoelectric heating rate and dust 
recombination cooling rate in the H~II region (Bakes
\& Tielens 1994), the temperature near the central star in the H~II
region rises, but the temperature near the gas shell hardly changes. 
Therefore, the existence of dust grains do not significantly alter
the results in this paper.

\subsection{Fragmentation of the Shell} 

The triggered star formation scenario predicts that the shocked layer
fragments and these fragments collapse to stars of the second 
generation. Although the fragmentation of the shocked layer is studied by
many authors, it is still uncertain which instability actually occurs
and which induces the next star formation. Here, we assume that
non-gravitational instabilities (e.g. decelerating shock instability
(DSI)) themselves are not the essential ones for the sequential star 
formation. 
Elmegreen (1989) showed that the gravitational instability couples with
DSI for the decelerating layer using the linear analysis, but 
MacLow \& Norman (1993) pointed out that the DSI quickly saturates 
in the non-linear phase. 

However, Garcia-Segura \& Franco (1996) showed the DSI is strongly
modified with the presence of the IF. The shell rapidly fragments and
the finger-like structures are formed in their 2D calculation.
Furthermore, the shadowing instability and corrugation instability of 
the IF were studied by Williams (1999, 2002) and these
instabilities also can disturb the IF and the shell. 
Although these complexities are beyond the scope of this paper,
it will be interesting to study the effect of these instabilities
on the evolution of the DF and the post-shock layer with 2D or 3D
calculations in the future.

Elmegreen \& Elmegreen (1978), 
Miyama, Narita \& Hayashi (1987), Lubow \& Pringle (1993),
and Nagai et al.(1998) have studied the stability of the dense gas layer 
and show that the earliest gravitational instability occurs with 
a wavelength which comparable to the layer thickness and 
the growth rate is $\sim ( G \rho )^{-1/2}$. 
The calculated shell suffers from this instability,
and the shaded region in Fig.2 represents the unstable region,
where $t \leq ( G \rho )^{-1/2}$. 
Although the mass scale at the maximum growth rate is comparatively
small, $\sim \sigma h^2 \sim 1~M_\odot$, larger mass scale perturbation
gradually grows in turn.  
As mentioned above, the gas
density is higher in the outer part of the shell, then the unstable
region extends from the outer edge to the inner part of the shell.
By the age of Sh104, 0.7~Myr, the outer part of the shell dominated
by CO molecule becomes unstable.
Therefore, the dense molecular fragments are formed around H~II
region and the triggered star formation may occur in the core of them,
as DV03 shows.

\subsection{The Role of this Triggering Process}

Even if the molecular gas is successfully gathered around the H~II
region and the triggered star formation actually occurs, its impact
on the global star formation is still uncertain. However, we can 
show its significance in a simple argument.
In our galaxy, most of stars are formed in clusters 
(e.g., Evans 1999), then the feedback effects from massive stars
are ubiquitous for the global star formation.
If every massive star gather the molecular gas of 
$M_{\rm sh}$ on average and the triggered
star formation occurs within the shell, the induced star formation
rate SFR$_{\rm HII}$ is estimated as
\begin{eqnarray}
\lefteqn{ {\rm SFR}_{\rm HII} \sim {\rm SFR}_0 }  \nonumber\\
&& \times
\Biggl(  \frac{M_{\rm av}}{0.6 M_\odot} \Biggr)^{-1}
\Biggl( \frac{f~(>20 M_\odot)}{0.0006} \Biggr)
\Biggl( \frac{M_{\rm sh}}{10^4 M_\odot} \Biggr)
\Biggl( \frac{\epsilon}{0.1} \Biggr)_,
\label{order}
\end{eqnarray}
where SFR$_0$ is the current total star formation rate in our galaxy, 
$M_{\rm av}$ is the average stellar mass, $f~(>20 M_\odot)$ is the 
number fraction of the massive star which creates relative large H~II 
regions $(>20M_\odot)$, and $\epsilon$ is the star formation efficiency
for the swept-up gas. 
The normalization values of $M_{\rm av} \sim 0.6 M_\odot$ and
$f~(>20 M_\odot) \sim 0.0006$ are calculated with the initial mass 
function (IMF) given by Miller \& Scalo (1979). Equation
(\ref{order}) shows that if the average mass of the gas shell
is $\sim 10^4 M_\odot$, this triggering process even alone can sustain 
the current galactic star formation rate.   
In our calculation, the molecular gas of $\sim 10^4 M_\odot$ 
is accumulated around one massive star owing to the expansion of the
H~II region by the time of $t \sim 1$~Myr. Therefore, this mode
of triggered star formation should be an important process.
The efficiency of triggering should depend on the physical conditions,
such as type of the central star, ambient density structure and so on.
For example, many H~II regions show the ``blister-like'' 
or ``champagne flow'' (Tenorio-Tagle 1979) features.
In these cases, the IF rapidly erodes the parental cloud and 
only a part of the swept-up mass remains within the shell 
, which might result in smaller $M_{\rm sh}$ 
(e.g., Whitworth 1979, Franco et al. 1994).
More realistic estimation including such cases will be explored in 
Paper II.

\section{Conclusion}

We have calculated the dynamical expansion of H~II region and
PDR around a massive star, focusing on the dense gas shell around
H~II region. Our results are simply summarized as follows,
\begin{itemize}
\item[1]
The dense gas shell is formed just in front of the IF, and this shell
is initially atomic but gradually dominated with molecules.
This is partly because the IF and SF overtake the preceding DFs
and SF enters the molecular region, and partly because
the reformation timescale within the shell becomes shorter than
the dynamical time because of the high density of the shell.
\item[2]
The calculation successfully reproduce the observed properties
of Sh104. The mass of the shell and H~II region,
and the location of the IF at $t \sim 0.7$~Myr agree with the observation.
The gravitational instability occurs after the gas shell
is dominated with the molecular gas. By the time of $t \sim 0.7$~Myr, 
the gravitationally unstable molecular shell is formed around H~II
region, as the observation shows.
\item[3]
In our calculation, the molecular gas shell of 
the order of $\sim 10^4 M_\odot$ is formed in 1~Myr around one massive
star. With simple estimation, we show the triggering process owing to
the expansion of H~II region is important in context of Galactic
star formation.  
\end{itemize}



\begin{thebibliography}{}

\bibitem{AK04}
Arthur, S.J., Kurtz, S.E., Franco, J. \& Albarran, Y. 2004, \apj, 608, 282


\bibitem{BT94}
Bakes, E.L.O. \& Tielens, A.G.G.M. 1994, \apj, 427, 822

\bibitem{Deh03} 
Deharveng, L. et al. 2003, \aap, 408, 25L

\bibitem[Diaz-Miller et al. (1998)]{DFS98}
Diaz-Miller, R.I., Franco, J. \& Shore, S.N. 1998, \apj, 501, 192

\bibitem{EL77}
Elmegreen, B.G. \& Lada, C.J. 1977, \apj, 214, 725

\bibitem{EE78}
Elmegreen, B.G. \& Elmegreen, D.M., 1978, \apj, 220, 1051

\bibitem{Elm89}
Elmegreen, B.G. 1989, \apj, 340, 786

\bibitem{Elm98}
Elmegreen, B.G., 1998, in Woodward, C.E., Shull, M., Thronson, H.A.,
eds, ASP Conf.Ser.Vol.148, Origins, Astron.Soc.Pac.San Francisco, p.150

\bibitem{Ev99}
Evans, N.J.II, 1999, \araa, 37, 311

\bibitem{FST94}
Franco, J., Shore, S.N. \& Tenorio-Tagle, G. 1994, \apj, 436, 795

\bibitem{FTB90} Franco, J., Tenorio-Tagle, G. \& Bodenheimer, P. 1990, \apj,
349, 126

\bibitem{GF96} Garcia-Segura, G. \& Franco, J.  1996, \apj, 469, 171


\bibitem{HM79} Hollenbach, D. \& McKee, C.F. 1979, \apjs,41, 555 

\bibitem{HM89} Hollenbach, D. \& McKee, C.F. 1989, \apj, 342, 306

\bibitem{HTT91} Hollenbach, D., Takahashi, T. \& Tielens,A.G.G.M. 
1991, \apj, 377, 192

\bibitem{hi04} Hosokawa, T. \& Inutsuka, S., 2005, in prep. (Paper II)

\bibitem{ki00}
Koyama, H. \& Inutsuka, S. 2000, \apj, 532, 980

\bibitem{Lk66}
Lasker, B.M. 1966, \apj, 143, 700
 
\bibitem{lp93}
Lubow, S. H. \& Pringle, J.E. 1993, \mnras, 263, 701

\bibitem{MN93}
Mac Low, M. \& Norman, M.L. 1993, \apj, 407, 207

\bibitem{M1965} Mathews,W.G. 1965, \apj, 142, 1120

\bibitem{MS79}
 Miller, G.E. \& Scalo, J.M. 1979, \apjs, 41, 513

\bibitem{MNH87}
 Miyama, S.M., Narita, S. \& Hayashi, C. 1987, Prog.Theor.Phys., 78, 105

\bibitem{NIM98}
 Nagai, T., Inutsuka, S. \& Miyama, S.M. 1998, \apj, 506, 306

\bibitem{NL97} Nelson, R.P. \& Langer, W.D. 1997, \apj, 482, 796

\bibitem{RM04}
Roger, R.S., McCutcheon, W.H., Purton, C.R. \& Dewdney, P.E. 2004, \aap,
	425, 553

\bibitem{RD92}
Roger, R.S. \& Dewdney, P.E. 1992, \apj, 385, 536

\bibitem{Sp78}
Spitzer, L. 1978, {\it Physical Processes in the Interstellar Medium}
(New York: Wiley).


\bibitem{Tt76} Tenorio-Tagle, G. 1976, \aap, 53, 411

\bibitem{Tt79} Tenorio-Tagle, G. 1979, \aap, 71, 59

\bibitem{vL97} van Leer, B. 1979, J.Comput.Phys., 32, 101

\bibitem{WD01} Weingartner, J.C. \& Draine, B.T. 2001, \apj, 548, 296

\bibitem{Wt79} Whitworth, A.P. 1979, \mnras, 186, 59

\bibitem{wt94}
Whitworth, A,P., Bhattal, A.S., Chapman, S.J., Disney, M.J., 
\& Turner, J.A. 1994, \aap, 268, 291 

\bibitem{Wl99}
 Williams, R.J.R. 1999, \mnras, 310, 789

\bibitem{Wl02}
 Williams, R.J.R. 2002, \mnras, 331, 693

\end{thebibliography}
\end{document}